\documentclass[letterpaper]{jpconf}
\usepackage{graphicx}
\begin{document}
\title{Precision Electroweak Measurements from CDF}

\author{Oliver Stelzer-Chilton}

\address{TRIUMF, Vancouver, Canada (on behalf of the CDF Collaboration)}

\begin{abstract}
We present results from electroweak precision measurements at CDF. These include the measurement of the $Z$ boson rapidity distribution using di-electron events, 
the $W$ boson charge asymmetry measurement in the electron-neutrino decay and the direct measurement of the $W$ boson mass and width in both the 
electron-neutrino and muon-neutrino decay channels.
\end{abstract}
\section{Introduction}
At the Tevatron, $W$ and $Z$ bosons are predominantly produced through quark-antiquark annihilation within the colliding proton and anti-proton beams. The bosons decay into two fermions which, 
due to their large mass, typically have a high momentum in the direction transverse to the beam. The $W$ and $Z$ boson decay channels, where the charged lepton is an electron or muon, have clean experimental 
signatures with low background contamination and can therefore be utilized for $W$ and $Z$ boson precision measurements. Events with a 
$Z$ boson\footnote{The exchange of a virtual photon and Z-$\gamma^*$ interference are indistinguishable from pure Z exchange. For the remainder of this document we use $Z$ boson to represent Z/$\gamma^*$, including interference.} are identified by the detection of two charged leptons, each with high transverse momentum. The electron energy is measured with the electromagnetic calorimeter and the muon momentum with the CDF 
central drift chamber. $W$ boson events are identified by one high transverse momentum charged lepton and one high transverse momentum neutrino. Since the neutrino does not interact with the detector, its existence is 
inferred by the transverse momentum imbalance in the detector. We describe four recent measurements from CDF. Two that constrain the parton distribution function of the proton,
the $W$ boson charge asymmetry and the $Z$ boson rapidity measurements and two $W$ boson precision measurements of the $W$ boson mass and width.
\section{W Boson Charge Asymmetry}
\begin{figure}[h]
\begin{minipage}{18pc}
\includegraphics[height=11pc,width=19pc]{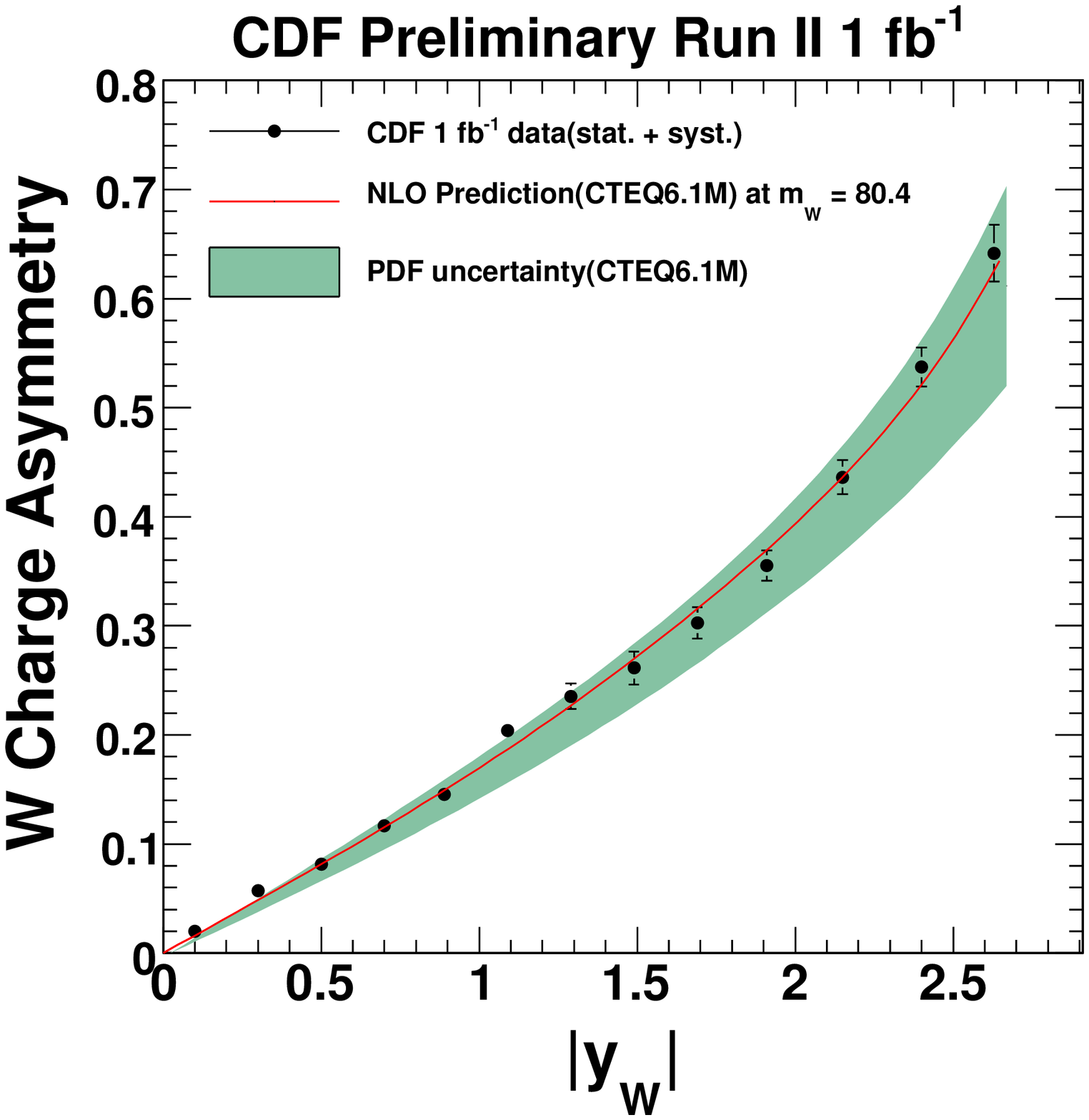}
\end{minipage}\hspace{1pc}%
\begin{minipage}{18pc}
\includegraphics[height=11pc,width=19pc]{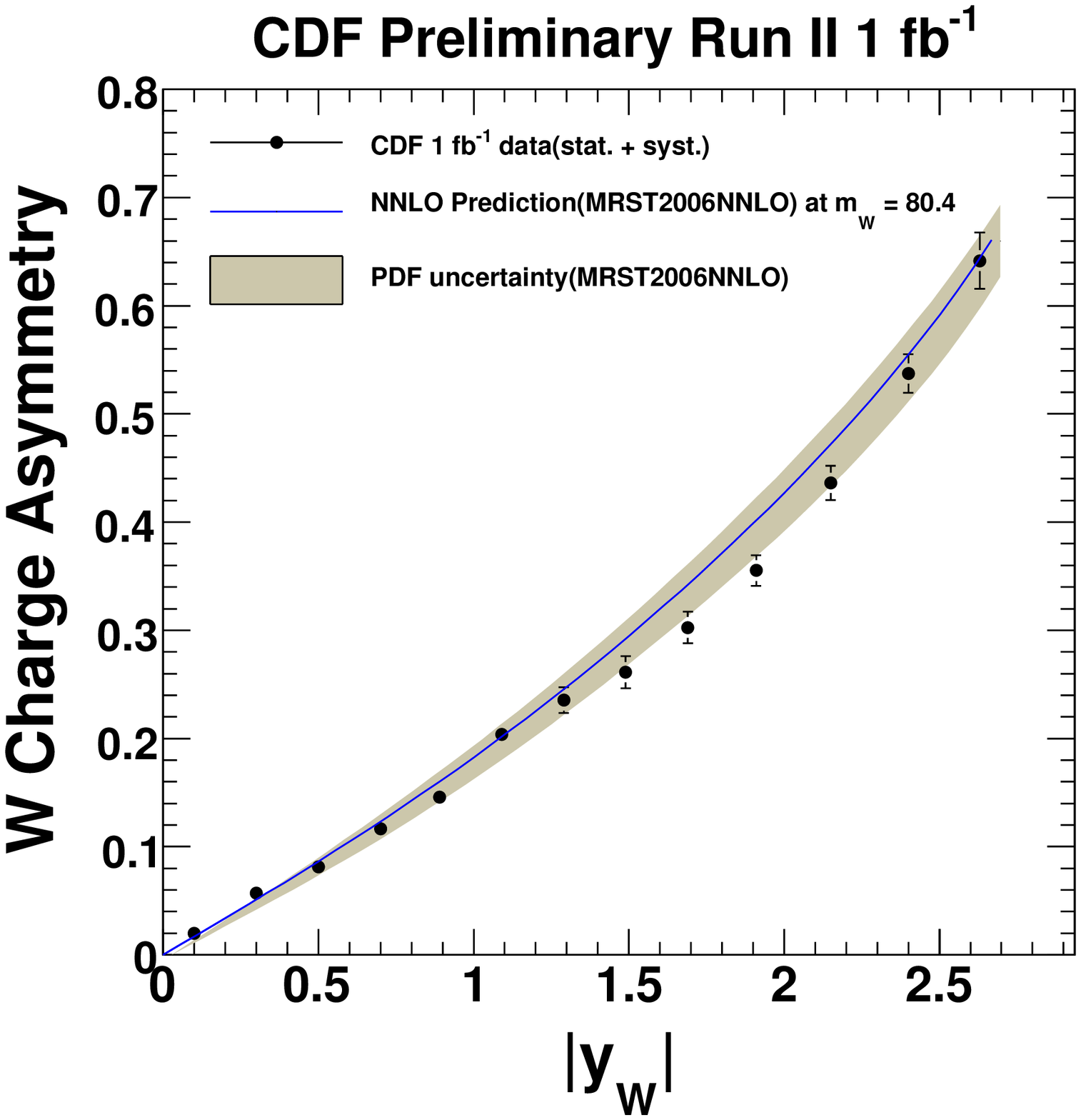}
\end{minipage}
\caption{The measured $W$ boson charge asymmetry and theoretical predictions from NLOCTEQ6.1 (left) and NNLO MRST 2006 (right), with their associated PDF 
uncertainties.}
\label{wcharge}
\end{figure}

On average the $u$($\bar{u}$) quark carries a higher fraction of the (anti)proton's momentum than the $d$($\bar{d}$) quark, thus
a $W^+$($W^-$) produced via a $u\bar{d}$($d\bar{u}$) annihilation will tend to be boosted along the (anti)proton beam.
This results in a non-zero forward-backward $W$ boson asymmetry, defined as:
\begin{equation}
A(y_W)=\frac{d\sigma(W^+)/dy_W-d\sigma(W^-)/dy_W}{d\sigma(W^+)/dy_W+d\sigma(W^-)/dy_W}
\end{equation}
where $y_W$ is the $W$ boson rapidity and $d\sigma/dy_W$ is the differential cross section for $W^+$ or $W^-$ boson production.
A measurement of $A(y_W)$ is sensitive to the ratio of $u$ and $d$ quark components of the protons parton distribution functions (PDF).
Since the longitudinal neutrino momentum is not known, a measurement of the electron or muon
charge asymmetry, rather than $A(y_W)$, has traditionally been made. This distribution is a convolution of $A(y_W)$
and the $V-A$ asymmetry from $W$ boson decays and the two asymmetries tend to cancel at large pseudorapidities.
In this analysis \cite{cdfcharge}, using 1 fb$^{-1}$ of $W\rightarrow e\nu$ data, $A(y_W)$ is extracted by constraining $M_W$ to its measured value, giving two possible solutions, each receiving a probability 
weight according to the decay structure and $\sigma (W^{\pm})$. The measured $W$ boson charge asymmetry is compared to the prediction from PDFs in Figure \ref{wcharge}
using  NLO CTEQ6.1M \cite{cteq} (left) and NNLO MRST 2006 \cite{mstw} (right).
The respective PDF uncertainties are shown with shaded bands. Given the small experimental uncertainties this measurement will help to constrain PDFs in future fits.
\section{Z Boson Rapidity}
$Z$ boson production at the Tevatron proceeds predominantly through annihilation of a $u$ (or $d$) quark in the proton and a $\bar{u}$ (or $\bar{d}$) quark 
in the anti-proton. The two partons carry momentum fractions $x_1$ and $x_2$. At leading order (LO), the momentum fraction $x_1$ ($x_2$) carried by the
two partons are related to the rapidity ($y$) of the $Z$ boson via the relation:
\begin{equation}
x_1=M_Ze^ys^{-1}, x_2=M_Ze^{-y}s^{-1}
\end{equation}
where $\sqrt{s}$ is the center of mass energy and $M_Z$ is the mass of the $Z$ bosons. Thus a measurement of $d\sigma(Z)/dy$ places constraints
on the proton PDFs. In particular, the high rapidity region probes both the high and low $x$ regions of the PDFs.
CDF has made a measurement of $d\sigma(Z)/dy$ using 2.1 fb$^{-1}$ of $Z\rightarrow ee$ data with $|$$\eta$$|$ $<$ 2.8.
This measurement has recently been included in fits by the MSTW collaboration \cite{mstwnew} and
\begin{figure}[b]
\begin{minipage}{18pc}
\includegraphics[height=10pc,width=19pc]{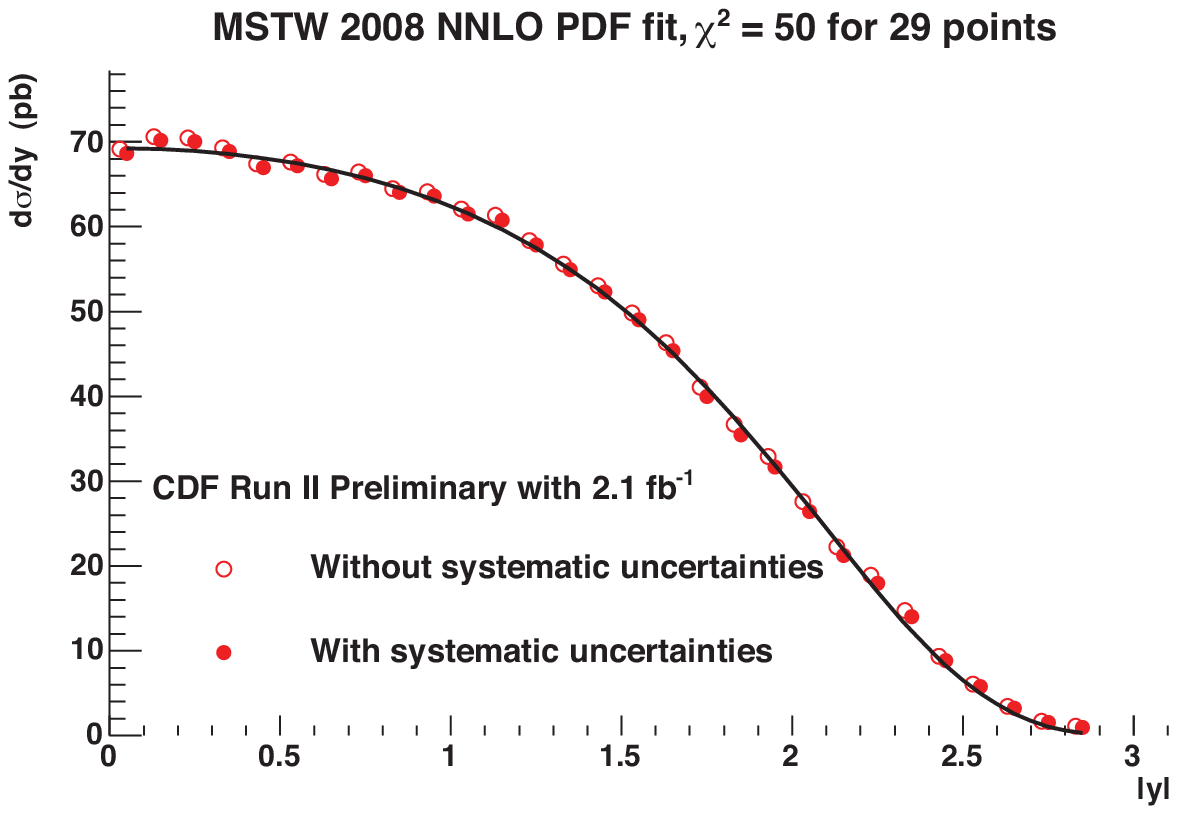}
\end{minipage}\hspace{1pc}%
\begin{minipage}{18pc}
\includegraphics[height=10pc,width=19pc]{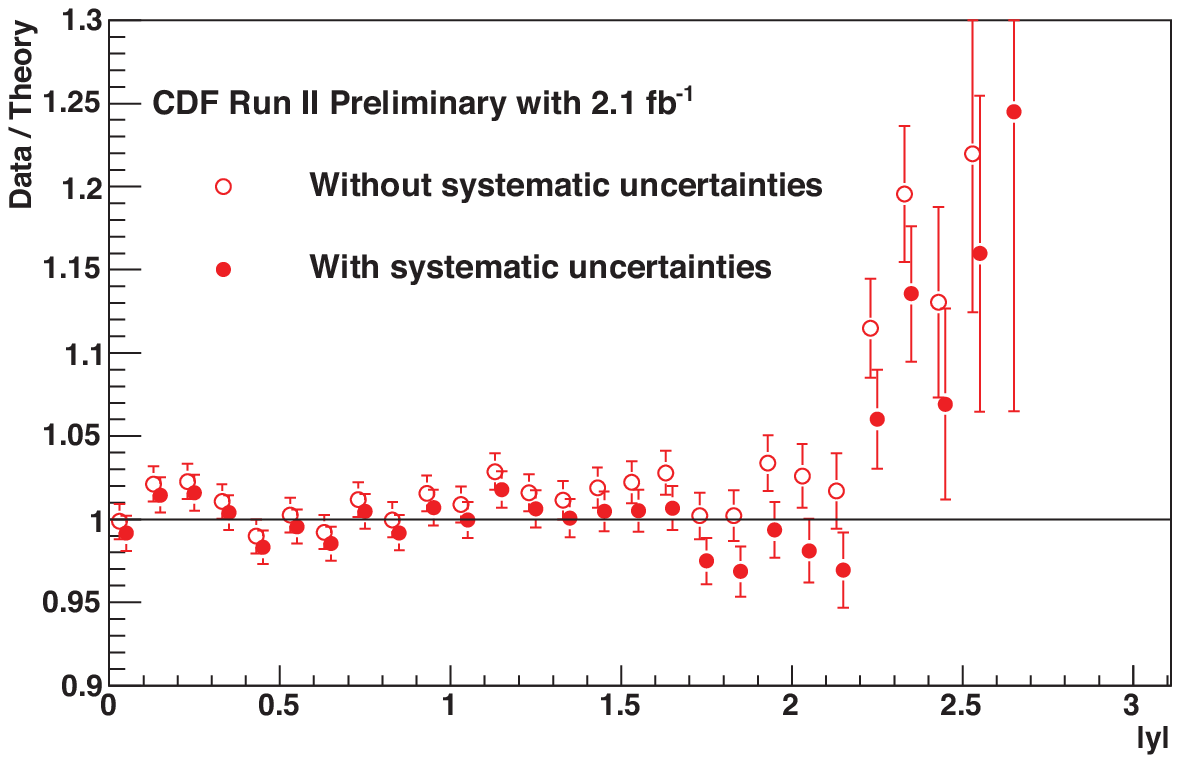}
\end{minipage}
\caption{The comparison of data and theory for the $Z$ boson rapidity distribution \cite{mstwnew}.}
\label{zrapidity}
\end{figure}
Figure \ref{zrapidity} shows the results of the NNLO fit.
The data are well fitted, particularly when the systematic uncertainties 
are taken into account. There is evidence of a slight excess of data over theory at high rapidity. 
\section{W Boson Mass Measurement}
\begin{figure}[h]
\begin{minipage}{18pc}
\includegraphics[height=11pc,width=19pc]{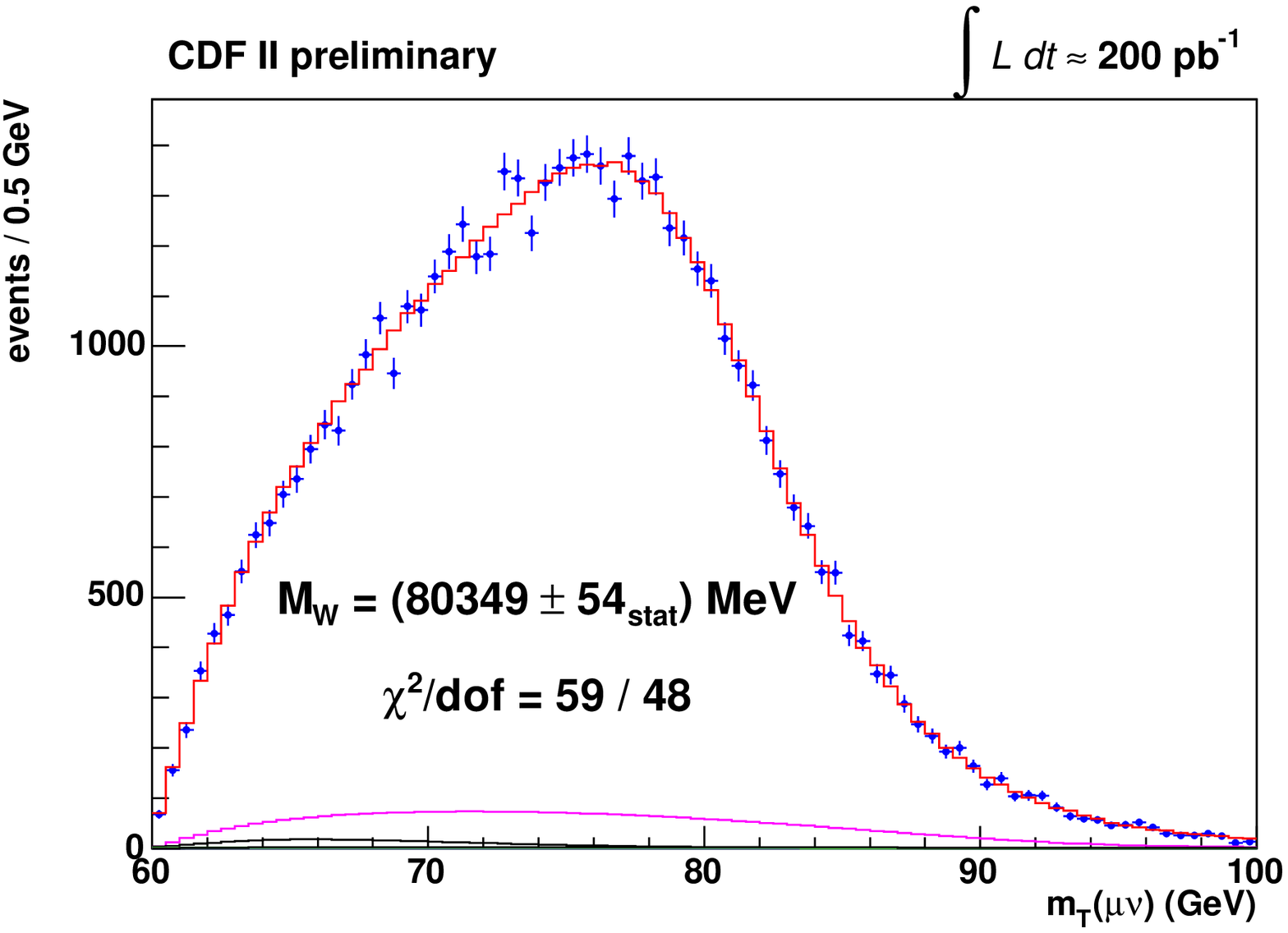}
\end{minipage}\hspace{1pc}
\begin{minipage}{18pc}
\includegraphics[height=11pc,width=19pc]{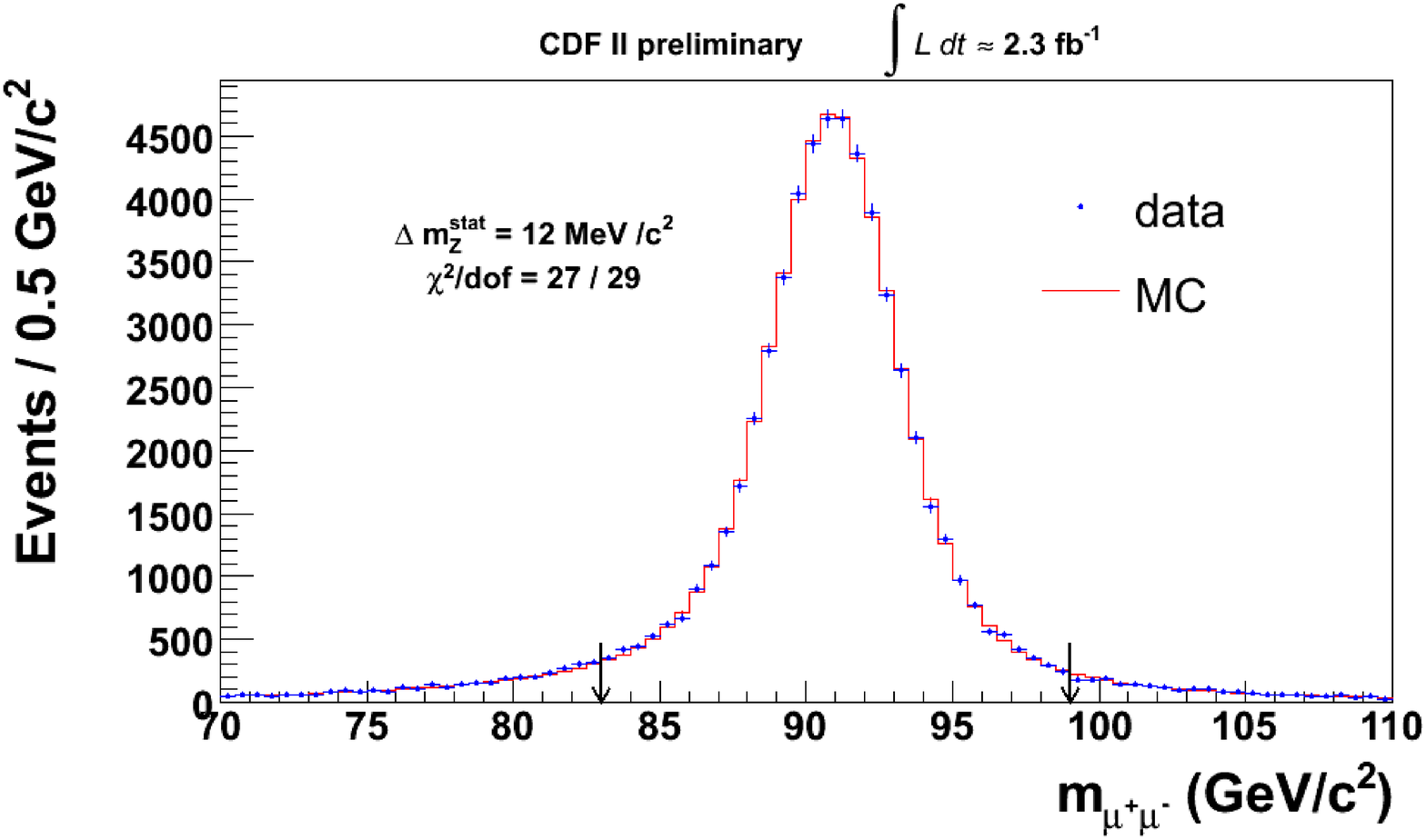}
\end{minipage}
\caption{Left: Transverse mass fit in the muon decay channel for the 200 pb$^{-1}$ measurement. Right: Invariant mass fit in the di-muon decay channel using 2.3 fb$^{^-1}$.}
\label{wmass}
\end{figure}

The $W$ boson mass receives self-energy corrections due to vacuum fluctuations involving virtual particles. Thus the $W$ boson 
mass probes the particle spectrum in nature, including particles that have yet 
to be observed directly. In order to constrain the Higgs boson mass, we need to account for the radiative corrections to $M_W$ due to the dominant top-bottom 
quark loop diagrams. For fixed values of 
other inputs, the current uncertainty on the top quark mass, $m_T$=173.1$\pm$1.3 GeV/$c^2$ \cite{top} corresponds to an 
uncertainty in its $W$ boson mass correction of 8 MeV/$c^2$. The current world average from direct measurements at LEP \cite{lepwmass} and Tevatron \cite{tevwmass}
yield a world average\footnote{Since the Institute took place, the D0 collaboration has released a $W$ boson mass measurement  
$m_W$ $=$ 80401$\pm$43 MeV/$c^2$ \cite{d0wmass}. The Gfitter collaboration subsequently computed a new world average of 
80399$\pm$23 MeV/$c^2$ \cite{gfitter}.} of 80398$\pm$25 MeV/$c^2$.  
It is clearly profitable to reduce the $W$ boson mass uncertainty further as a means of constraining the Higgs boson mass. The current one-sided 95\% 
C.L. upper limit from indirect electroweak constraints is $m_H$$<$163 GeV/c$^2$ \cite{higgs}.
At the Tevatron, the $W$ boson mass is extracted from a template fit to the Jacobian edge of the transverse mass distribution, defined as $m_T=\sqrt{2 p_T^l
p_T^{\nu}(1-cos[\phi^l-\phi^{\nu}])}$, where $p_T^l$ is the transverse momentum of the charged lepton and $p_T^{\nu}$ is the inferred neutrino transverse momentum.
A fast Monte Carlo simulation is used to model the lineshape of the template distributions, accounting for the detector response, resolution and acceptance
effects. We constrain these important detector and physics effects by control samples and calculations which determine the 
precision of the measurement of $m_W$. Examples of the effects that need to be rigorously studied, extracted and modeled, include the tracker momentum scale and resolution, 
the calorimeter energy response, scale and resolution, the intrinsic $W$ boson transverse momentum, internal QED radiation and the proton parton distribution functions (PDFs).
Using 200 pb$^{-1}$ of data, CDF combines the measurement in the muon and electron decay channel and obtains $m_W$=80413$\pm$34(stat)$\pm$34(syst) MeV/$c^2$ 
\cite{wmass1} \cite{wmass2}.
Figure \ref{wmass} on the left shows the transverse mass fit for the muon decay channel. 
Many of the systematic uncertainties in this measurement scale with the statistics of the control samples used to calibrate the detector and can therefore be improved 
with an analysis of the larger datasets in hand. Additionally, improvements in the production and decay model such as improved PDF fits are likely to further reduce the overall systematic 
uncertainty on future measurements of $m_W$. CDF has begun analyzing an over ten times larger data sample with an integrated luminosity of 2.3 fb$^{-1}$, as can be seen in the $Z$ boson mass fit on the right
side of Figure \ref{wmass}. Studies in progress confirm that many of the systematic uncertainties scale with luminosity as expected and we look forward to an 
updated $m_W$ measurement with a precision better than the current world average of 25 MeV/c$^2$.
\section{W Boson Width Measurement}
\begin{figure}[h]
\begin{minipage}{18pc}
\includegraphics[height=11pc,width=19pc]{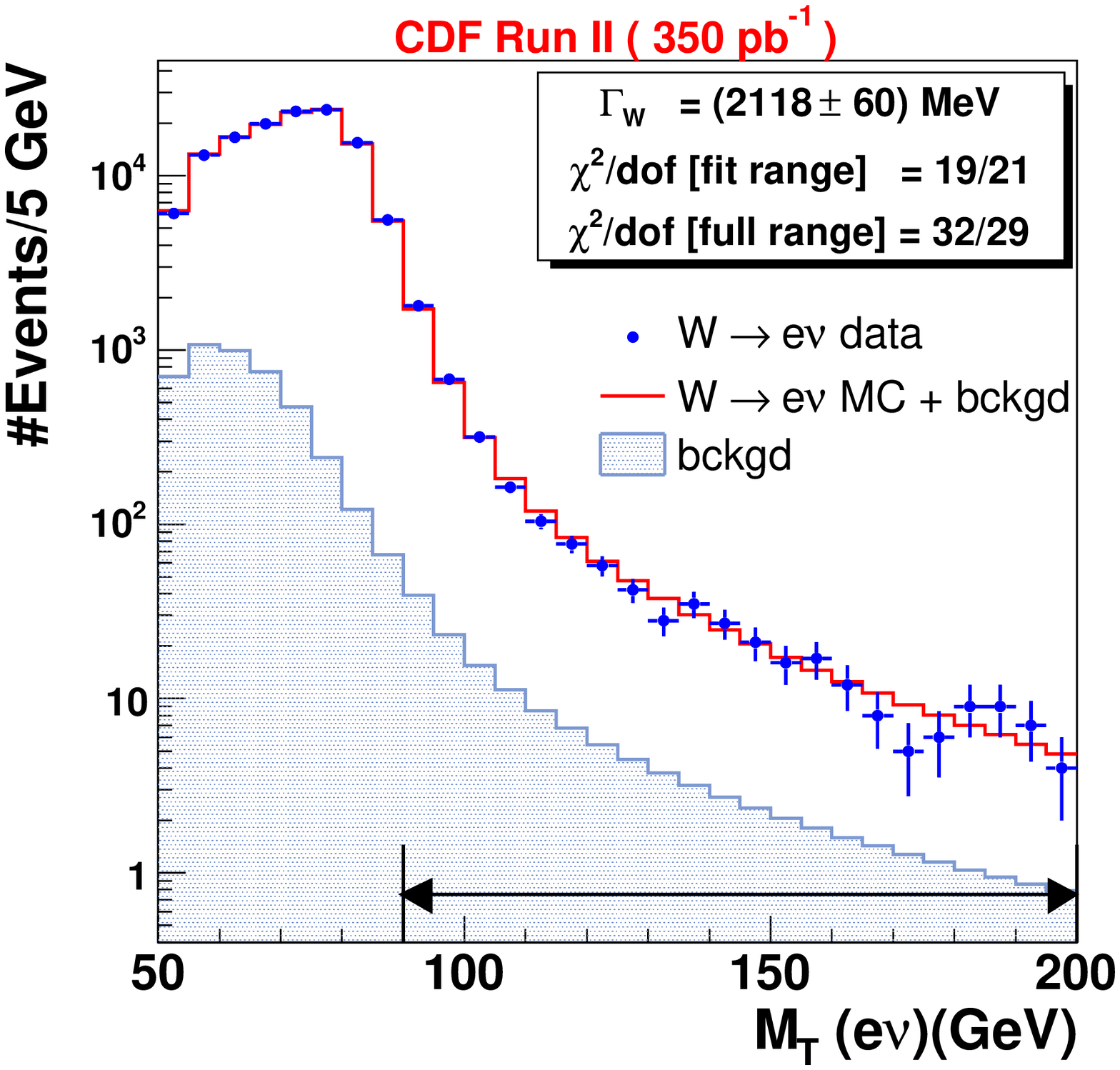}
\end{minipage}\hspace{1pc}
\begin{minipage}{18pc}
\includegraphics[height=11pc,width=19pc]{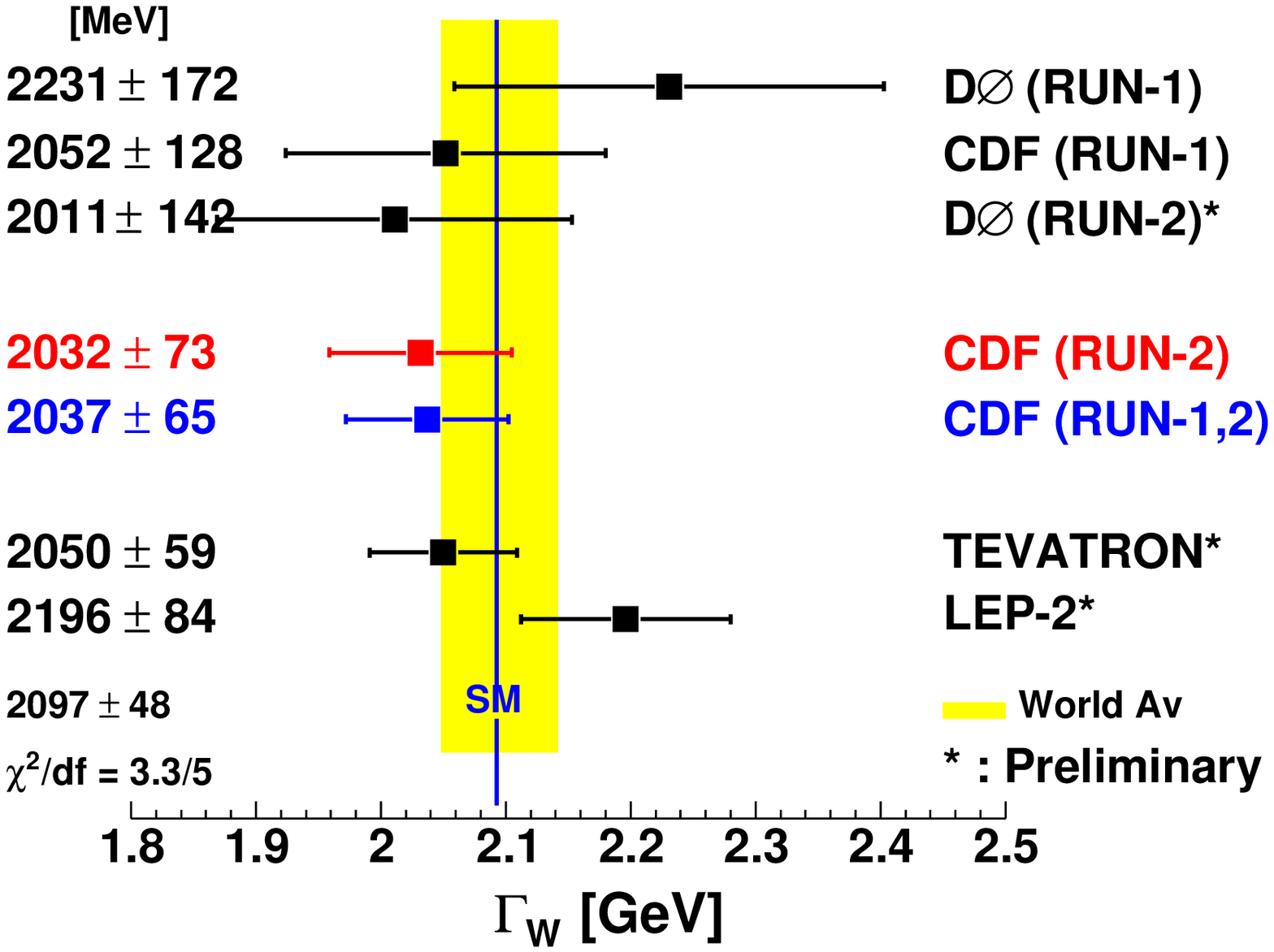}
\end{minipage}
\caption{Left: Transverse mass fit in the electron decay channel for the 350 pb$^{-1}$ measurement. Right: Comparison with other Tevatron
measurements and LEP-2.}
\label{wwidth}
\end{figure}

The $W$ boson width is known within the Standard Model to an extremely high precision of 0.1\%, thus an accurate
experimental measurement is desirable to test this prediction.
The measurement technique is very similar to the extraction of the $W$ boson mass, but instead of fitting the Jacobian edge, the fit is performed
in the tail region 90 $<$ $M_T$ $<$ 200 GeV/c$^2$ of the transverse mass distribution, where the Gaussian detector resolution falls off more quickly compared 
to the $W$ boson Breit-Wigner lineshape. CDF uses 350 $pb^{-1}$ analyzing the electron and muon decay channels to measure the $W$ boson width \cite{wwidth1}.
Figure \ref{wwidth} shows the transverse mass fits for $W\rightarrow e\nu$ events.
The results of the muon and electron decay channel are combined to give
the final result $\Gamma_W$ = 2032 $\pm$ 73 MeV/c$^2$, the world's most precise measurement, which is in good agreement with the
Standard Model prediction of 2091 $\pm$ 2 MeV/c$^2$ \cite{wwidthsm}.
\section{Conclusion}
The large Tevatron datasets of $W$ and $Z$ bosons with their clean decays to leptonic final states allow for a rich 
electroweak physics program. The CDF $W$ boson charge asymmetry and $Z$ boson rapidity measurements improve the constraints
on proton PDFs, which in turn are important inputs to precision measurements of the $W$ boson mass and $W$ boson width. 
The anticipated more precise determination of the $W$ boson mass from CDF is essential for a further constraint
of the Higgs boson mass and will test the Standard Model with higher precision.
\newline\newline
I would like to thank my CDF collaborators, especially those whose work went into the results presented here at the beautiful 
Lake Louise Winter Institute.
\newline

\end{document}